\documentclass[12pt]{article}

\usepackage{graphicx}
\usepackage{amssymb}
\usepackage{amsmath}
\begin{document}

\begin{center}
{\Large{\bf Morphological Phase Separation in Unstable Thin Films: 
Pattern Formation and Growth}} \\
\ \\
\ \\
by \\
Prabhat K. Jaiswal$^1$, Manish Vashishtha$^2$, Sanjay Puri$^{1}$, and Rajesh Khanna$^2$ \\
$^1$School of Physical Sciences, Jawaharlal Nehru University, \\
New Delhi -- 110067, India. \\
$^2$Department of Chemical Engineering, Indian Institute of Technology Delhi,
New Delhi -- 110016, India.
\end{center}

\begin{abstract}
We present results from a comprehensive numerical study of {\it morphological phase separation} 
(MPS) in unstable thin liquid films on a $2$-dimensional substrate. 
We study the quantitative properties of the evolution 
morphology via several experimentally relevant markers, e.g., correlation function, structure 
factor, domain-size and defect-size probability distributions, and growth laws. 
Our results suggest that the late-stage morphologies exhibit dynamical scaling, and their 
evolution is self-similar in time. We emphasize the analogies and differences between MPS in films
and segregation kinetics in unstable binary mixtures.
\end{abstract}

\newpage

\section*{Introduction}

Nonequilibrium processes are of great importance in science and technology. The
properties of a system are governed by both its constituents and the kinetic 
processes of formation. In this context, there has been intense research interest in the {\it kinetics 
of phase transitions}, i.e., the nonequilibrium evolution of a system which has been rendered 
thermodynamically unstable by a sudden change of parameters, e.g., temperature, pressure, etc. 
\cite{pw09}. 

Two important problems in this area are the {\it kinetics of unstable thin films} 
\cite{dg85,odb97,bjm09,cm09,sk98,ksr00} and {\it phase-separation kinetics in unstable mixtures} 
\cite{sp09,ab94,bf91,ao02}. We have recently initiated a study of the analogies and differences
between these two physical problems \cite{kav10, vjk10}. Both systems are characterized by
the emergence and growth of domains. In unstable films, the coarsening domains consist of flat
regions with the equilibrium height $h_m$. These are separated by high-curvature regions with
gradients in the height 
field. For $h_m = 0$, we have {\it true dewetting} (TD) \cite{kas09}, where holes are punctured in the 
film. For $h_m > 0$, the evolution is referred to as {\it morphological phase separation} (MPS) 
\cite{sj93,js94,sr02}. On the other hand, in phase-separating binary ($AB$) mixtures, the
growing domains consist of coexisting $A$-rich and $B$-rich regions, which are locally in
equilibrium. In both systems, coarsening is governed by transport of material
via diffusive and hydrodynamic processes. The underlying growth mechanisms dictate the
{\it system morphology} and {\it domain growth laws}.

In this communication, we present results from a comprehensive numerical study of MPS in 
3-dimensional thin films. We use several experimentally relevant quantities to completely 
characterize the emergent morphology. The quantities we measure (e.g., correlation function, 
structure factor, domain-size distributions) have often been used to describe evolution 
morphologies in segregating mixtures. However, to the best of our knowledge, these have 
not been used in the context of unstable films. At appropriate places in this paper, we 
will highlight the novel features of MPS in thin films.

\section*{Model and Simulation Details}

The evolution of unstable films is modeled by a continuity equation for the height field.
This is obtained by simplifying the hydrodynamic equations of 
motion via the {\it lubrication approximation} \cite{rj74}. In dimensionless units \cite{kav10}, 
the continuity equation has the form: 
\begin{equation}
 \frac{\partial}{\partial T} H(\vec{X},T) = \vec{\nabla}\cdot 
 \left[M \vec{\nabla} \left(\frac{\delta F}{\delta H}\right)\right],
 \label{eq1:ch}
\end{equation}
where $H(\vec{X},T)$ denotes the film height at space point $\vec{X}$ (lying on a $d=2$ substrate) 
and time $T$. The unstable initial state of the film is $H(\vec{X},0)=1+$ small fluctuations, 
i.e., the film is homogeneous to start with. In our dimensionless rescaling, all heights are 
measured in units of the (dimensional) initial film height $h_0$. 
In Eq.~(\ref{eq1:ch}), the height-dependent mobility $M(H) = H^3$, corresponding 
to Stokes flow with no slip. The equilibrium states are determined from the free-energy 
functional:
\begin{equation}
 F[H] = \int d\vec{X} \left[ f(H) + 
 \frac{1}{2}{\left(\vec{\nabla}H\right)}^2\right],
 \label{eq2:f}
\end{equation}
where $f(H)$ is the local free energy, and the square-gradient term measures the surface 
tension. We consider a thin film on a coated substrate with a {\it long-range van der Waals
attraction} due to the substrate, and a {\it short-range van der Waals repulsion} due to the
coating \cite{kjs96}. The corresponding dimensionless potential is
\begin{equation}
 f(H)=-\frac{1}{6}\left[ \frac{1-R}{\left(H + D\right)^{2}}+ \frac{R}{H^{2}}\right].
 \label{eq3:f-tf}
\end{equation}
In Eq.~(\ref{eq3:f-tf}), $R$ is the ratio of the effective Hamaker constants, and $D$ is 
the dimensionless coating thickness in units of $h_0$.

\begin{figure}[!htb]
\centering
\includegraphics*[width=0.6\textwidth]{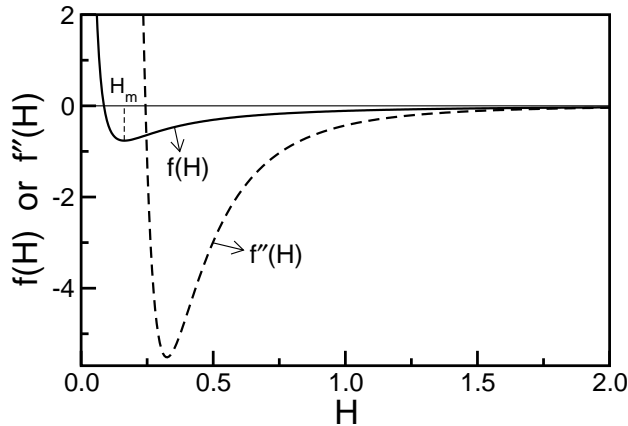}
\caption{Plot of the free energy $f(H)$ in Eq.~(\ref{eq3:f-tf}), and 
its second derivative $f^{\prime\prime}(H)$, with film thickness $H$. The 
parameter values are $R=-0.1$ and $D=0.2$. The thickness of the equilibrium 
flat film phase is given by $H_m=D|R|^{1/3}/[(1+|R|)^{1/3} - |R|^{1/3}]$.
For $f^{\prime\prime}(H) < 0$, the homogeneous thin film is spontaneously unstable
and segregates into flat domains and high-curvature droplets.}
\label{fig:fig1}
\end{figure}

In Fig.~\ref{fig:fig1}, we plot $f(H)$ vs. $H$ for typical parameter values, $R=-0.1$ and 
$D=0.2$. We also plot $f^{\prime\prime} (H)$ vs. $H$ in Fig.~\ref{fig:fig1}. The homogeneous 
thin film is spontaneously unstable to fluctuations about the initial state ($H_0 = 1$) 
when $f^{\prime\prime} (1) < 0$. Replacing 
Eqs.~(\ref{eq2:f})-(\ref{eq3:f-tf}) in Eq.~(\ref{eq1:ch}), we obtain the nonlinear evolution 
equation:
\begin{eqnarray}
 \frac{\partial H}{\partial T} &=& \vec{\nabla}\cdot
 \left[H^3\left(f^\prime(H)-\nabla^2 H \right) \right] \nonumber \\
 &=& \vec{\nabla}\cdot\left[H^3\vec{\nabla}
 \left(\frac{1-R}{3(H+D)^3} + \frac{R}{3H^3} - \nabla^2 H \right) \right].
 \label{eq4:h}
\end{eqnarray}
The linear stability analysis of Eq.~(\ref{eq4:h}) shows that the most unstable mode has 
a wavelength
\begin{equation}
 L_M = \frac{4 \pi}{\sqrt{-f^{\prime\prime} (1)}} = 4\pi
 \left[R + \frac{1-R}{(1+D)^4}\right]^{-1/2}.
 \label{eq5:Lm}
\end{equation}

The final states of the evolution are obtained from the double-tangent construction for 
$f(H)$ in Fig.~\ref{fig:fig1}. The film segregates into phases with 
$H=H_m=D|R|^{1/3}/[(1+|R|)^{1/3} - |R|^{1/3}]$ and $H=\infty$. The latter value results 
in steepening high-curvature regions which never reach ``equilibrium''. It is relevant 
to ask whether Eq.~(\ref{eq4:h}) has a static {\it bump} solution $H(X)$, with 
$H(X)\rightarrow H_m$ as $X\rightarrow \pm \infty$. Such a solution must satisfy the 
zero-current condition:
\begin{equation}
 \frac{d^2H}{dX^2} - f^\prime (H) = c,
 \label{eq6:d2h}
\end{equation}
where the constant $c=0$ as the bump is flat when $X\rightarrow \pm \infty$, i.e.,
$d^2H/dX^2 \rightarrow 0$. The first integral of Eq.~(\ref{eq6:d2h}) yields
\begin{equation}
 {\left(\frac{dH}{dX}\right)}^2 = 2\left[ f(H) - f(H_m)\right].
 \label{eq7:dh2}
\end{equation}
As $H=H_m$ is the sole minimum of $f(H)$ (see Fig.~\ref{fig:fig1}), the right-hand-side of 
Eq.~(\ref{eq7:dh2}) can only be zero at $H=H_m$. This immediately rules out a bump solution,
which must satisfy $dH/dX = 0$ at its peak position. The above scenario should be 
contrasted with the phase separation of $AB$ mixtures, where the system segregates into 
coexisting $A$-rich and $B$-rich domains. As we discuss shortly, this fundamental difference 
will have important consequences for the evolution morphology in thin films.

We numerically solve Eq.~(\ref{eq4:h}) in $d=3$ (i.e., on a $d=2$ substrate),
starting with a small-amplitude ($\simeq 0.01$) random perturbation about the mean film thickness
$H = 1$. The parameters $D$ and $R$ were chosen so that the film is spinodally unstable at $H = 1$.
The system size is $V=(32 L_{M})^2$. Periodic boundary conditions are applied at the lateral ends.
A 32-point grid per $L_{M}$ was found to be sufficient when central differencing in space with half-node 
interpolation was combined with {\it Gear's algorithm} for time-marching. This scheme is especially 
suitable for stiff equations. 

\section*{Numerical Results and Discussion}

\begin{figure}[!htb]
\centering
\includegraphics*[width=0.6\textwidth]{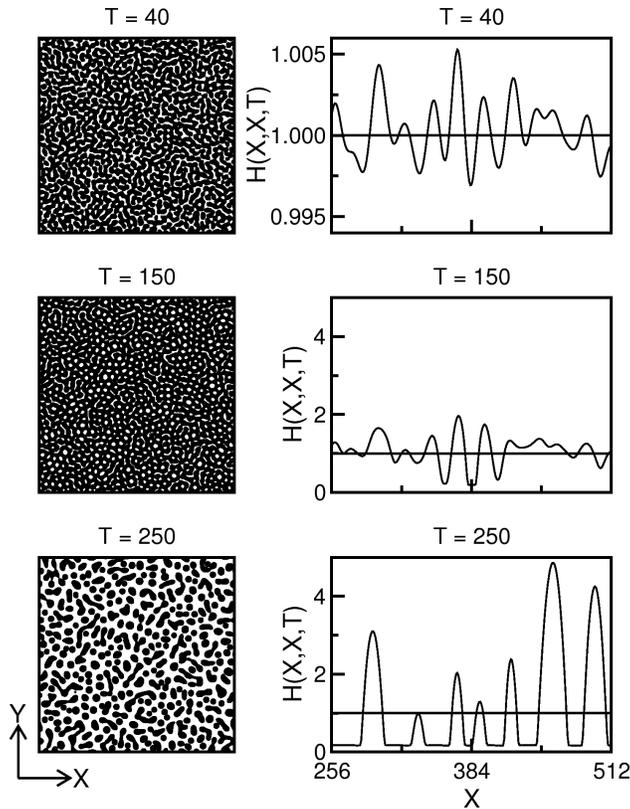}
\caption{Kinetics of {\it morphological phase separation} (MPS) in an unstable 
thin film in $d=3$. The initial condition for the evolution consisted of uniformly-distributed 
small-amplitude random fluctuations about $H(X,Y,T=0)=1$. The parameter values were $R=-0.1$ and 
$D=0.2$. Other simulation details are provided in the text. The dimensionless evolution times
are specified above each frame. Left: Snapshots of the height field. Regions with $H>1$
(high-curvature phase) are marked in black, whereas those with $H<1$ (flat phase) are unmarked. 
Right: Variation of the height field along a diagonal cross-section ($Y=X$) of the snapshots.} 
\label{fig:fig2}
\end{figure}

In Fig.~\ref{fig:fig2}, we show MPS in a thin film evolving from the unstable homogeneous 
state. The parameter values are $R=-0.1$ and $D=0.2$. 
The snapshots on the left show the height field: regions with $H<1$ 
are unmarked, and regions with $H>1$ are marked black. The conservation law ensures that
$\bar{H} = \int d\vec{X} H(\vec{X},T)/V$ is constant in time: $\bar{H}=H_0=1$ in this case.
The frames on the right show the variation of the height field along a diagonal cross-section 
of the snapshots [$H(X,Y=X,T)$ vs. $X$]. The early time regime ($T=40$) corresponds to the 
growth of fluctuations about the homogeneous state. This growth is exponential and can be 
obtained from a linear stability analysis of Eq.~(\ref{eq4:h}) \cite{jvk}: $H(\vec{X},T) =
1+ \delta H(\vec{X},T)$, where
\begin{equation}
\delta H(\vec{X},T) \simeq \exp \left\{t\nabla^2 [f''(1)-\nabla^2] \right\} \delta H(\vec{X},0).
\end{equation}

The growing fluctuations are saturated by the nonlinearity for $H\simeq H_m$ ($T=150$): there is 
no corresponding saturation for the regions with $H>1$. In the late stages ($T=250$), there is 
growth of domains with $H=H_m$ (flat phase). At the same time, the defects (or hills) become 
sharper and sharper. Domain growth is driven by the transport of liquid from smaller hills 
to larger hills due to the chemical-potential gradient. This transport can be {\it hydrodynamic} 
(when the black regions are connected, e.g., $T=150$) or {\it diffusive} (when the black regions 
are not connected, e.g., $T=250$). 

\begin{figure}[!htb]
\centering
\includegraphics*[width=0.6\textwidth]{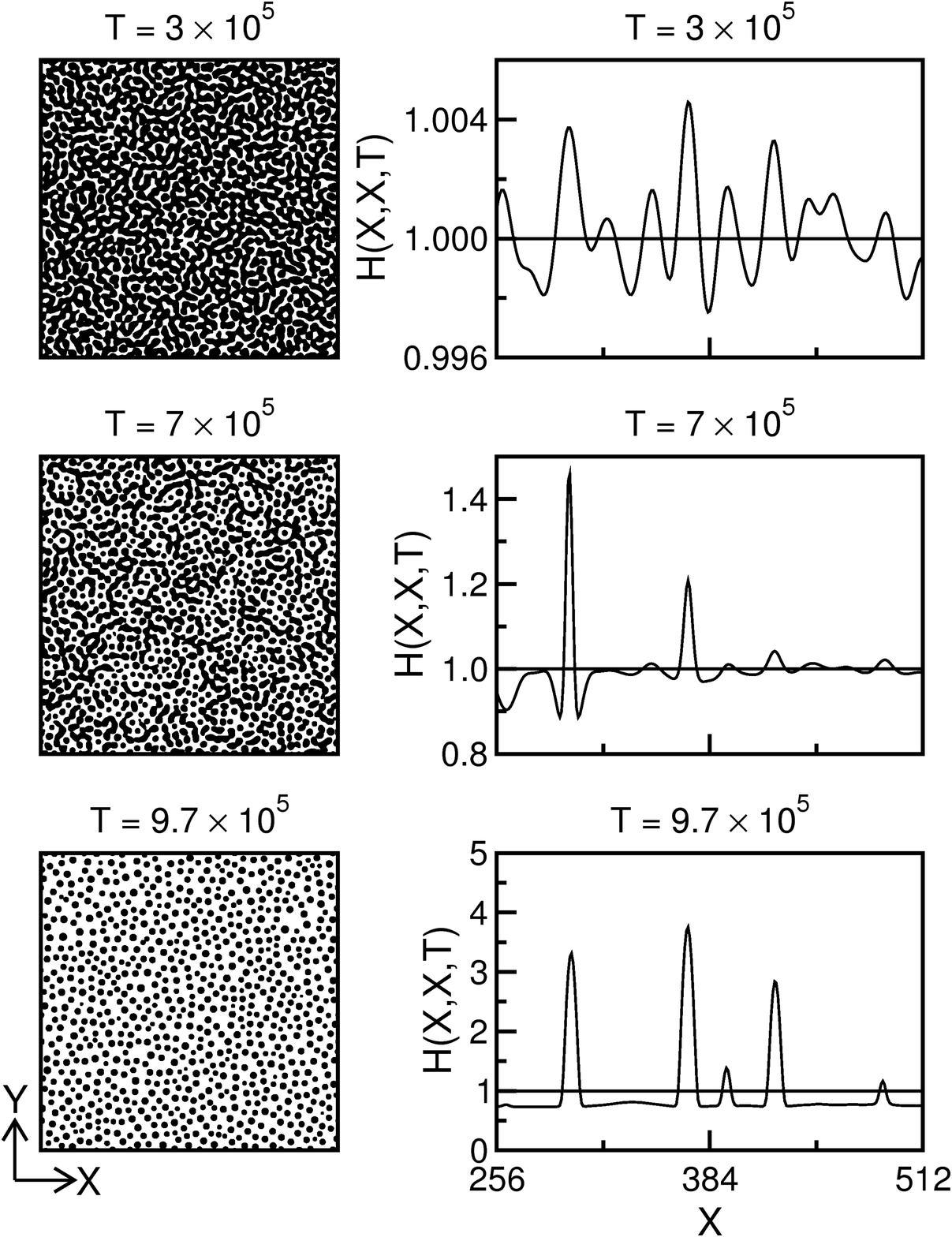}
\caption{Analogous to Fig.~\ref{fig:fig2}, but for the parameter values $R=-0.1$ and $D=0.8$ 
(corresponding to a much thicker film).}
\label{fig:fig3}
\end{figure}

Notice that the morphology evolves continuously in Fig.~\ref{fig:fig2}: 
from {\it bicontinuous} ($T=40$) to {\it circular domains of flat phase} ($T=150$) to {\it droplets 
of high-curvature phase} ($T=250$). We expect that the system enters a {\it scaling regime} for 
$T \gtrsim 250$, where the morphology becomes self-similar and only the domain size grows. 
A similar evolution is seen for much thicker films with $D=0.8$ in 
Fig.~\ref{fig:fig3}. As expected, the time-scales of MPS are much longer for the thick film. 
The statistical results we show subsequently correspond to the film with $D=0.2$, whose 
evolution is shown in Fig.~\ref{fig:fig2}. 

Let us present some simple arguments to understand the evolution of the scaling morphology in 
Fig.~\ref{fig:fig2} (e.g., $T=250$) and Fig.~\ref{fig:fig3} (e.g., $T=9.7 \times 10^5$). Let $L(T)$ and
$H_d(T)$ denote the characteristic 
size and height of a droplet of the high-curvature phase at time $T$. If the number of droplets in 
the system is $N(T)$, the amount of surplus liquid in the defects $\sim NL^2H_d$. 
The corresponding depletion of liquid in the flat domains is $(V-NL^2)(1-H_m)$, where 
$V$ is the substrate area. As there is a conservation of liquid, we have
\begin{equation}
 NL^2H_d \simeq (V-NL^2)(1-H_m),
 \label{eq8:n}
\end{equation}
or
\begin{equation}
 \phi(T) \simeq \frac{(1-H_m)}{L^2(H_d + 1 -H_m)} \simeq \frac{(1-H_m)}{L^2 H_d}.
 \label{eq9:phi}
\end{equation}
In Eq.~(\ref{eq9:phi}), $\phi(T)=N(T)/V$ denotes the number of droplets per unit area. In the final
step of Eq.~(\ref{eq9:phi}), we have neglected $(1-H_m)$ (which is constant) with respect to
$H_d$ (which grows).

How do these droplets grow with time? Their coarsening proceeds via an 
{\it evaporation-condensation} mechanism. The smaller droplets evaporate and their 
material is diffusively transported to the larger droplets through the flat domains. 
The chemical potential at a point is estimated as
\begin{equation}
 \mu = f^\prime (H) - \nabla^2 H.
 \label{eq10:mu}
\end{equation}
We assume that the droplets have a conical profile (cf. Fig.~\ref{fig:fig2}) with the height
\begin{equation}
 H(r) \simeq H_d - (H_d-1) \left(\frac{r}{L}\right)^n,
 \label{eq11:hr}
\end{equation}
where the droplet is centered at $r=0$, and $n$ is the profile exponent. Then
\begin{equation}
 \nabla^2 H = \frac{1}{r}\frac{\partial}{\partial r}\left(r\frac{\partial H}{\partial r}
 \right) \simeq -\frac{\left(H_d-1\right)n^2}{rL}.
 \label{eq12:d2h}
\end{equation}
Thus, we estimate the chemical potential at the droplet boundary ($H=1$) as 
\begin{equation}
 \mu \simeq f^\prime (1) + \frac{(H_d-1)n^2}{L^2} \simeq f^\prime (1) + \frac{H_d n^2}{L^2}.
 \label{eq13:mu2}
\end{equation}
A better estimate of the chemical potential can be obtained by integrating over the droplet. 
The resultant value is the same as that in Eq.~(\ref{eq13:mu2}) up to geometric prefactors.

The chemical-potential gradient occurs over flat domains (with $H=H_m$) of size $\sim L$, and the 
corresponding current is
\begin{eqnarray}
 J \simeq -M(H_m) |\vec{\nabla} \mu| \simeq \frac{H_m^3 H_d n^2}{L^3}.
 \label{eq14:J}
\end{eqnarray}
Thus, the droplets grow as
\begin{equation}
 \frac{dL}{dT} \sim \frac{H_m^3 H_d n^2}{L^3}.
 \label{eq15:dl}
\end{equation}
In the scaling regime, we expect $H_d \sim L$ so that $L(T) \sim {(H_m^3 T)}^{1/3}$. This 
growth law is well known in the context of phase-separation kinetics, and is referred to 
as the {\it Lifshitz-Slyozov} (LS) growth law \cite{pw09, op87}. In earlier work \cite{kav10, vjk10}, 
we have demonstrated that MPS in the $d=2$ film (i.e., on a $d=1$ substrate) also obeys 
the LS growth law. The above arguments apply for $d=2$ films also, except when $n=1$
in Eq.~(\ref{eq11:hr}).

There are several experimental tools to quantitatively characterize the MPS evolution 
morphologies. The {\it correlation function} of the height field is defined as
\begin{equation}
 C(\vec{r},T) \equiv C(r,T) = \langle \delta H(\vec{X},T) \delta H(\vec{X}+\vec{r},T)\rangle, 
 \label{eq16:C}
\end{equation}
where $\delta H$ ($= H-H_0$) denotes the fluctuation in the height field. The angular brackets 
in Eq.~(\ref{eq16:C}) denote an averaging over independent runs. The statistical results 
presented here are obtained as an average over $10$ runs. The correlation function depends only on 
the distance $r$ (the magnitude of $\vec{r}$) because the system is translationally invariant 
and isotropic. In scattering experiments with, e.g., light, X-rays, neutrons, etc., we measure the 
{\it structure factor}, which is the Fourier transform of $C(\vec{r},T)$ with wave-vector 
$\vec{k}$:
\begin{equation}
 S(\vec{k},T)\equiv S(k,T)= \int d\vec{r}\, e^{i \vec{k}\cdot\vec{r}} C(\vec{r},T).
 \label{eq17:S}
\end{equation}

\begin{figure}[!htb]
\centering
\includegraphics*[width=0.52\textwidth]{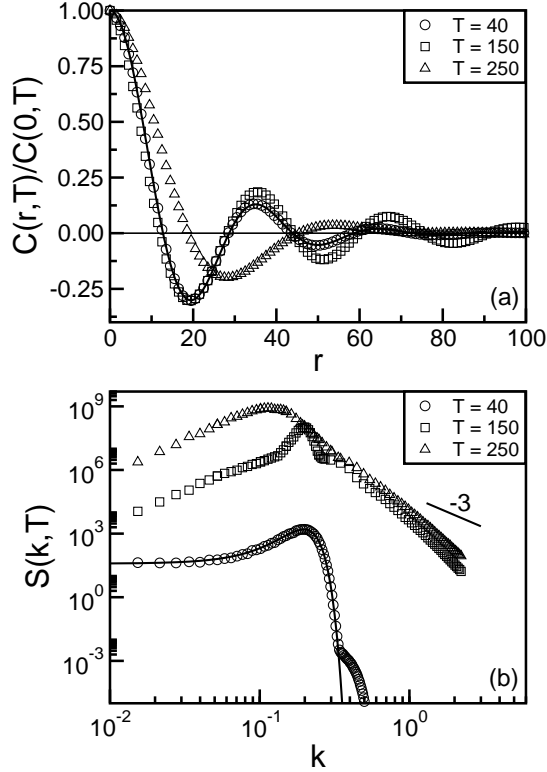}
\caption{ Statistical quantities for the evolution depicted in Fig.~\ref{fig:fig2}. 
These are obtained on $(32 L_M)^2$ systems as an average over $10$ independent runs. 
(a) Plot of the spherically-averaged correlation function $C(r,T)/C(0,T)$ vs. $r$. 
The solid line denotes the result from linear theory in Eq.~(\ref{eq19:Cl}), evaluated at $T=40$. 
(b) Plot of the spherically-averaged structure factor $S(k,T)$ vs. $k$. 
The solid line denotes the expression in Eq.~(\ref{eq18:Sl}), evaluated at $T=40$.
The line with slope $-3$ refers to Porod's law in $d=2$, $S(k,T) \sim k^{-(d+1)}$ for large $k$. 
The Porod law characterizes scattering off sharp interfaces \cite{porod,op87}. 
In the unstable thin film, there are no equilibrium interfaces as there is only 
one homogeneous phase.}
\label{fig:fig4}
\end{figure}

Let us present numerical results for the statistical quantities defined above. 
In Fig.~\ref{fig:fig4}(a), we plot the spherically-averaged correlation function 
[$C(r,T)/C(0,T)$ vs. $r$] for the evolution depicted in Fig.~\ref{fig:fig2}. 
At early times, linear theory applies and the corresponding structure factor is 
\cite{jvk} 
\begin{equation}
 S_\textrm{lin}(k,T) \simeq AV \exp \left[2k^2\left(\alpha - k^2\right) T\right],
 \label{eq18:Sl}
\end{equation}
where $A$ is the amplitude of initial fluctuations, $\langle \delta H (\vec{X},0)
\delta H (\vec{X^\prime},0)\rangle = A\delta (\vec{X}-\vec{X^\prime})$. In Eq.~(\ref{eq18:Sl}), $V$ 
denotes the system volume, and $\alpha = -f^{\prime\prime}(1)$. The early-time correlation 
function is obtained as the inverse Fourier transform of $S_\textrm{lin}(k,T)$:
\begin{eqnarray}
 C_\textrm{lin}(r,T)&=& AV\int \frac{d\vec{k}}{(2\pi)^2} e^{-i\vec{k}\cdot\vec{r}} 
 \exp \left[2k^2\left(\alpha - k^2\right)T\right] \nonumber \\
 &=& \frac{AV}{2\pi}\int_0^\infty dk \, kJ_0(kr) 
 \exp \left[2k^2\left(\alpha - k^2\right)T\right].
 \label{eq19:Cl}
\end{eqnarray}

The solid line in Fig.~\ref{fig:fig4}(a) denotes the expression for $C_\textrm{lin}(r,T)/C_\textrm{lin}(0,T)$ 
from Eq.~(\ref{eq19:Cl}) with $T=40$. It is in excellent agreement with the numerical data for 
$T=40$. As time goes on, the fluctuations grow exponentially and linear theory ceases to
hold. There is no change 
in the length scale during the exponential growth regime -- hence, the zero-crossings of the 
correlation function at $T=150$ in Fig.~\ref{fig:fig4}(a) are comparable to those for $T=40$.
(The length-scale data shown 
later demonstrates that domain growth occurs for $T \gtrsim 150$.) There is a crossover 
in $C(r,T)$, as expected from the morphological evolution seen in Fig.~\ref{fig:fig2}. The 
snapshot at $T=250$ is indicative of the scaling morphology, viz., droplets of the 
high-curvature phase in a background of the flat phase. In the late stages, we expect 
{\it dynamical scaling} of $C(r,T)$ and $S(k,T)$ \cite{bs74, pw09}: 
\begin{eqnarray}
 C(r,T) &=& g(r/L), \nonumber \\
 S(k,T) &=& L^d f(kL),
 \label{eq20:CS}
\end{eqnarray}
where $L$ is the characteristic length scale. In Eq.~(\ref{eq20:CS}), $g(x)$ and $f(p)$ are 
scaling functions which do not depend on time. In Fig.~\ref{fig:fig4}(a), the correlation 
function at $T=250$ lies in the scaling regime.

In Fig.~\ref{fig:fig4}(b), we plot the spherically-averaged structure factor 
[$S(k,T)$ vs. $k$, on a log-log scale] for the same times as in Fig.~\ref{fig:fig4}(a). 
As before, there is a crossover in the functional form of the structure factor. 
The data set for $T=40$ corresponds to the linear regime, and is in good agreement 
with the expression in Eq.~(\ref{eq18:Sl}). Notice that the linear theory is not 
valid at very large values of $k$ ($\gtrsim 0.5$) where we see effects of discreteness 
of the simulation lattice. The data set at $T=250$ corresponds to the asymptotic scaling 
form in Eq.~(\ref{eq20:CS}). The peak location [$k_m(T)\sim L^{-1}$] moves to smaller values 
as the characteristic length scale grows. There is a shoulder at higher values of 
$k$, which reflects the second length scale apparent in the 
snapshots in Fig.~\ref{fig:fig2}. This shoulder disappears as we approach the 
asymptotic droplet morphology. Finally, the tail of the structure factor does 
not show the Porod tail \cite{porod}, $S(k,T)\sim k^{-(d+1)}$ at large $k$, which is 
characteristic of phase-separating systems \cite{op87}. 
The Porod tail arises from scattering off sharp interfaces -- however, the MPS morphology in 
Fig.~\ref{fig:fig2} does not have any equilibrium interfaces at all. 

\begin{figure}[!htb]
\centering
\includegraphics*[width=0.52\textwidth]{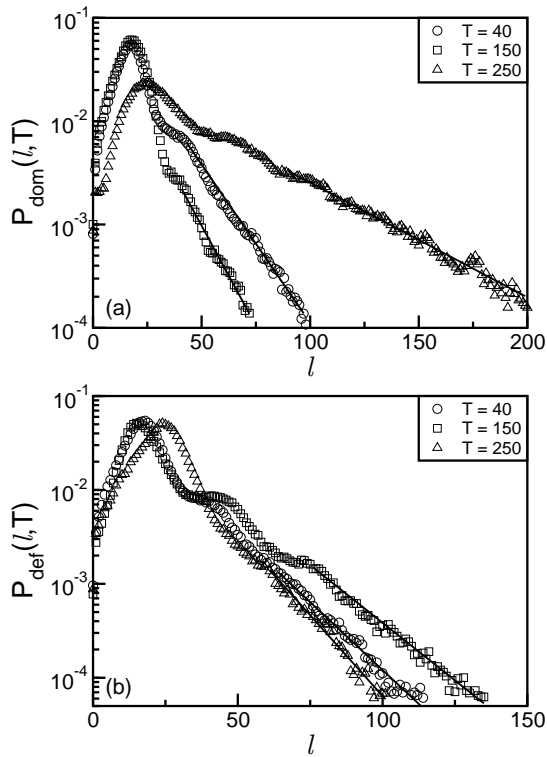}
\caption{Probability distributions of domain and defect sizes for the evolution depicted in 
Fig.~\ref{fig:fig2}. The data is plotted on a linear-log scale. The solid lines denote 
exponential fits to the tail region. 
(a) Domain-size distribution, $P_\textrm{dom}(l,T)$ vs. $l$. 
(b) Defect-size distribution, $P_\textrm{def}(l,T)$ vs. $l$.}
\label{fig:fig5}
\end{figure}

An alternative method of characterizing the growth morphology is via the probability 
distribution $P(l,T)$ of domain size $l$. Consider a typical snapshot in Fig.~\ref{fig:fig2}. 
We examine the variation of the height field in both the $x$- and $y$-directions, and keep 
track of the ``zero''-crossings (where $H=1$, the average height). This yields the distribution 
of sizes for the flat domains, as well as the defects or hills. In Fig.~\ref{fig:fig5}(a), 
we plot the domain-size distribution $P_\textrm{dom}(l,T)$ vs. $l$. This quantity also shows 
the expected crossover from early to late times. The data set for $T=40$ has a broader 
distribution than the data set for $T=150$, though the peak positions are similar. The 
$T=40$ morphology has a broader distribution
because it still contains modes with various length scales -- these are 
dominated by the most unstable wavelength ($L_M$) as the fluctuations grow. The peak positions 
at $T=40,150$ are comparable because there is no domain growth in the early exponential regime.
At later times, the domains coarsen and the peak position of $P_\textrm{dom}(l,T)$ shifts to
the right [see data set for $T=250$ in Fig.~\ref{fig:fig5}(a)]. The functional form at $T=250$
corresponds to the asymptotic scaling function $p_\textrm{dom}(x)$, which is defined as 
\begin{equation}
 P_\textrm{dom}(l,T)=L^{-1} p_\textrm{dom}(l/L).
 \label{eq21:P}
\end{equation}

We make two further observations about $P_\textrm{dom}(l,T)$. First, the plots at early
times ($T=40, 150$) show a marked shoulder, consistent with our observation of a 
two-scale morphology in the snapshots of Fig.~\ref{fig:fig2}, and the structure factors of 
Fig.~\ref{fig:fig4}(b). This shoulder disappears as 
we enter the asymptotic scaling regime. Second, the tail of the distribution decays exponentially,
i.e., linearly on the semi-log plot in Fig.~\ref{fig:fig5}(a). This is typical of morphologies with 
well-defined characteristic scales, and has been observed earlier in the context of phase-separation 
morphologies \cite{sp}.

In Fig.~\ref{fig:fig5}(b), we plot the defect-size distribution
[$P_\textrm{def}(l,T)$ vs. $l$] at the same times as in Fig.~\ref{fig:fig5}(a). 
This plot shows the same general features as Fig.~\ref{fig:fig5}(a).

\begin{figure}[!htb]
\centering
\includegraphics*[width=0.5\textwidth]{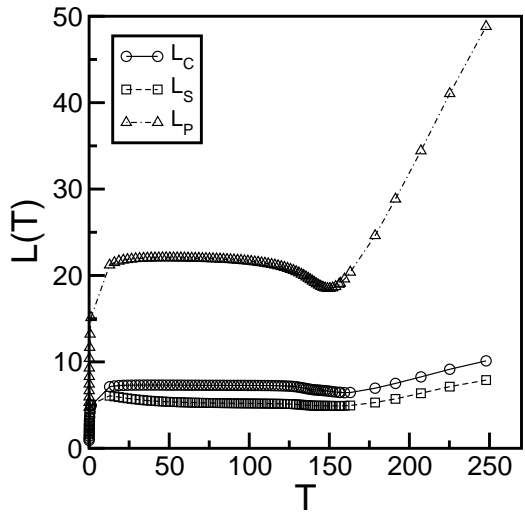}
\caption{Time-dependence of the characteristic length scale for the 
evolution in Fig.~\ref{fig:fig2}. We show data for three 
measures of the length scale: $L_C, L_S$, and $L_P$, obtained from 
$C(r,T), S(k,T)$, and $P_{\textrm{dom}}(l,T)$, respectively.}
\label{fig:fig6}
\end{figure}

Finally, in Fig.~\ref{fig:fig6}, we study the growth of the characteristic length scale. We use
three different measures of the length scale, obtained from the statistical quantities introduced
earlier:\\
(a) $L_{C}$, the length scale up to which $C(r,T)$ decays to half its maximum value 
(which arises at $r=0$).\\
(b) $L_{S}$, the inverse of the first moment of $S(k,T)$: $L_{S}=
{\langle k \rangle}^{-1} $ with  
\begin{equation}
 \langle k \rangle =\frac{\int_0^\infty{dk \, k S(k,T)}}{\int_0^\infty dk \, S(k,T)}.
 \label{eq22:k}
\end{equation}
(c) $L_{P}$, the average domain size from the relevant probability distribution: 
\begin{equation}
 L_{P}=\int_0^\infty dl \, l P_{\textrm{dom}}(l,T).
 \label{eq23:lp}
\end{equation}
We show the time-dependence of these length scales in Fig.~\ref{fig:fig6}. 
It is interesting to note that all
the different length scales show the same physical behavior. There is an {\it early regime} 
(up to $T\simeq 150$), where the initial fluctuations grow exponentially but there 
is almost no change in the length scale. In the context of phase-separation kinetics, this
is referred to as the {\it Cahn-Hilliard-Cook} (CHC) regime. This is followed by an {\it intermediate
regime}, where nonlinear effects saturate the growing fluctuations. The system segregates 
into domains of the flat phase and defects of the high-curvature phase. In the {\it late} stages 
($T>175$), there is growth of domains. It is not possible to ascertain the asymptotic growth 
exponent from our present results, which only access a small window in the scaling regime. 
For this, we require 1-2 further decades of growth, which would be computationally very expensive.
The numerical difficulty in accessing the late stages can be understood as follows. Recall that
the interfaces become progressively steeper with time. Therefore, we need a corresponding reduction
of the mesh size in space (and time) to resolve the interfaces properly.

\section*{Summary}

Let us conclude this paper with a brief summary. We have undertaken a comprehensive numerical
study of {\it morphological phase separation} (MPS) in unstable liquid films. In particular, we have used
several experimentally relevant tools to quantitatively characterize the emergent morphologies.
The statistical quantities we measure are the correlation function, structure factor, domain-size and 
defect-size probability distributions, and the corresponding growth laws. The study of these properties
provides a complete picture of the evolution dynamics. We make several important
predictions in this context. At late times, the system should enter a scaling regime, 
where the above quantities show {\it dynamical scaling}, i.e., the morphology becomes 
self-similar in time. We hope that our numerical study will motivate fresh experiments 
on unstable thin films.

There are important analogies and differences between MPS in thin films and phase-separation 
kinetics in binary mixtures. We have highlighted some of these features in this paper.
A crucial difference between the two systems is the nature of the coarsening domains and defects.
The unstable film segregates into droplets of a high-curvature phase (defects), separated by domains
of a flat phase. On the other hand, the unstable $AB$ mixture segregates into {\it coexisting} domains
of $A$-rich and $B$-rich phases, which are separated by near-equilibrium interface defects. This
difference has important consequences for the evolution morphologies and their quantitative properties.      

R.K. acknowledges the financial support of the Department of Science and Technology, India.

\end{document}